\documentclass[]{elsarticle}

\usepackage{lineno,hyperref}
\usepackage{amsmath}
\modulolinenumbers[5]
\usepackage{subfigure}

\journal{Journal}

\graphicspath{{figs/}}

\begin{document}

\begin{frontmatter}

\title{Fungi anaesthesia}

\author[1]{Andrew Adamatzky}
\author[2]{Antoni Gandia\footnote{At the time of experiments AG was affiliated with Mogu S.r.l., Inarzo, Italy}}

\address[1]{Unconventional Computing Laboratory, UWE, Bristol, UK\\\url{andrew.adamatzky@uwe.ac.uk}}
\address[2]{Institute for Plant Molecular and Cell Biology, CSIC-UPV, Valencia, ES}

\begin{abstract}
Electrical activity of fungus \emph{Pleurotus ostreatus} is characterised by slow (hours) irregular waves of baseline potential drift and fast (minutes) action potential likes spikes of the electrical potential. An exposure of the mycelium colonised substrate to a chloroform vapour lead to several fold decrease of the baseline potential waves and increase of their duration. The chloroform vapour also causes either complete cessation of spiking activity or substantial reduction of the spiking frequency.  Removal of the chloroform vapour from the growth containers leads to a gradual restoration of the mycelium electrical activity.
\end{abstract}


\end{frontmatter}


\section{Introduction}

Most living cells are sensitive to anaesthetics~\cite{sonner2008hypothesis,eckenhoff2008can,gremiaux2014plant}. 
First experiments on anaesthesia of plants have be done by Claude Bernard in late 1800s~\cite{bernard1974lectures}. Later experiments on amoeba~\cite{hiller1927action} shown that weak concentration of narcotics causes the amoebae to spread out and propagate in a spread condition while narcotic concentrations led to cessation of movements. During last century the experimental evidences mounted up including anaesthesia of yeasts~\cite{sonner2008hypothesis}, various aquatic invertebrates~\cite{oliver1991sensitivity}, plants~\cite{milne1999inhalational,gremiaux2014plant}, protists~\cite{nunn1974effect}, 
bronchial ciliated cells~\cite{verra1990effects}. A general consensus now is that any living substrate can be anaesthetised~\cite{gremiaux2014plant}. The question remains, however, how exactly species without a nervous system would respond to exposure to anaesthetics. 

In present paper we focus on fungi anaesthesia. Why fungi? The fungi is a largest, widely distributed and the oldest  group of living organisms~\cite{carlile2001fungi}. Smallest fungi are microscopic single cells. The largest (15 hectares) mycelium belongs to \emph{Armillaria gallica} (synonymous with \emph{A. bulbosa}, \emph{A. inflata}, and \emph{A. lutea}) ~\cite{smith1992fungus} and the largest fruit body belongs to \emph{Phellinus ellipsoideus} (formerly \emph{Fomitiporia ellipsoidea}) which weighs half-a-ton~\cite{dai2011fomitiporia}. 

Fungi exhibit a high degree of protocognitive abilities. For example, they are capable for efficient exploration of confined spaces~\cite{hanson2006fungi, held2008examining, held2009fungal, held2010microfluidics,held2011probing}. Moreover, optimisation of the mycelial network~\cite{boddy2009saprotrophic} is similar to that of the slime mould \emph{Physarum polycephalum}~\cite{adamatzky2009developing} and transport networks~\cite{adamatzky2012bioevaluation}. Therefore, we can speculate that the fungi can solve the same range of computational problems as \emph{P. polycephalum}~\cite{adamatzkyAdvancesPhysarum}, including shortest path~\cite{nakagaki2000intelligence,nakagaki2001smart,nakagaki2001path, nakagaki2007minimum,tero2010rules}, Voronoi diagram~\cite{shirakawa2009simultaneous}, Delaunay triangulation, proximity graphs and spanning tree, concave hull and, possibly, convex hull, and, with some experimental efforts, travelling salesman problem~\cite{jones2014computation}. The fungi's protocognitive abilities and computational potential make them fruitful substrates for anaesthesia because they might show us how non-neuron awareness is changing under effects of narcotics. 

We use extracellular electrical potential of mycelium as indicator of the fungi activity. Action potential-like spikes of electrical potential have been discovered using intra-cellular recording of mycelium of \emph{Neurospora crassa}~\cite{slayman1976action} and further confirmed in intra-cellular recordings of action potential in hyphae of \emph{Pleurotus ostreatus} and \emph{A. gallica}~\cite{olsson1995action} and in extra-cellular recordings of fruit bodies of and substrates colonized by mycelium of \emph{P. ostreatus}~\cite{adamatzky2018spiking}. While the exact nature of the travelling spikes remains uncertain we can speculate, by drawing analogies with oscillations of electrical potential of slime mould~\emph{Physarum polycephalum}~\cite{iwamura1949correlations, kamiya1950bioelectric, kishimoto1958rhythmicity, meyer1979studies}, that the spikes in fungi are triggered by calcium waves, reversing of cytoplasmic flow, translocation of nutrients and metabolites. Studies of electrical activity of higher plants can brings us even more clues. Thus, the plants use the electrical spikes for a long-distance communication aimed to coordinate the activity of their bodies~\cite{trebacz2006electrical,fromm2007electrical,zimmermann2013electrical}. The spikes of electrical potential in plants relate to a motor activity~\cite{simons1981role,fromm1991control,sibaoka1991rapid,volkov2010mimosa}, responses to changes in temperature~\cite{minorsky1989temperature}, osmotic environment~\cite{volkov2000green} and mechanical stimulation~\cite{roblin1985analysis,pickard1973action}.

The paper is structured as follows. We present the experimental setup in Sect.~\ref{methods}. Analysis of the electrical activity of the intact and anaesthetised fungi is given in Sect.~\ref{results}. Section~\ref{discussion} present some critique and directions for further research.

\section{Methods}
\label{methods}

\begin{figure}[!tbp]
    \centering
\subfigure[]{\includegraphics[width=0.49\textwidth]{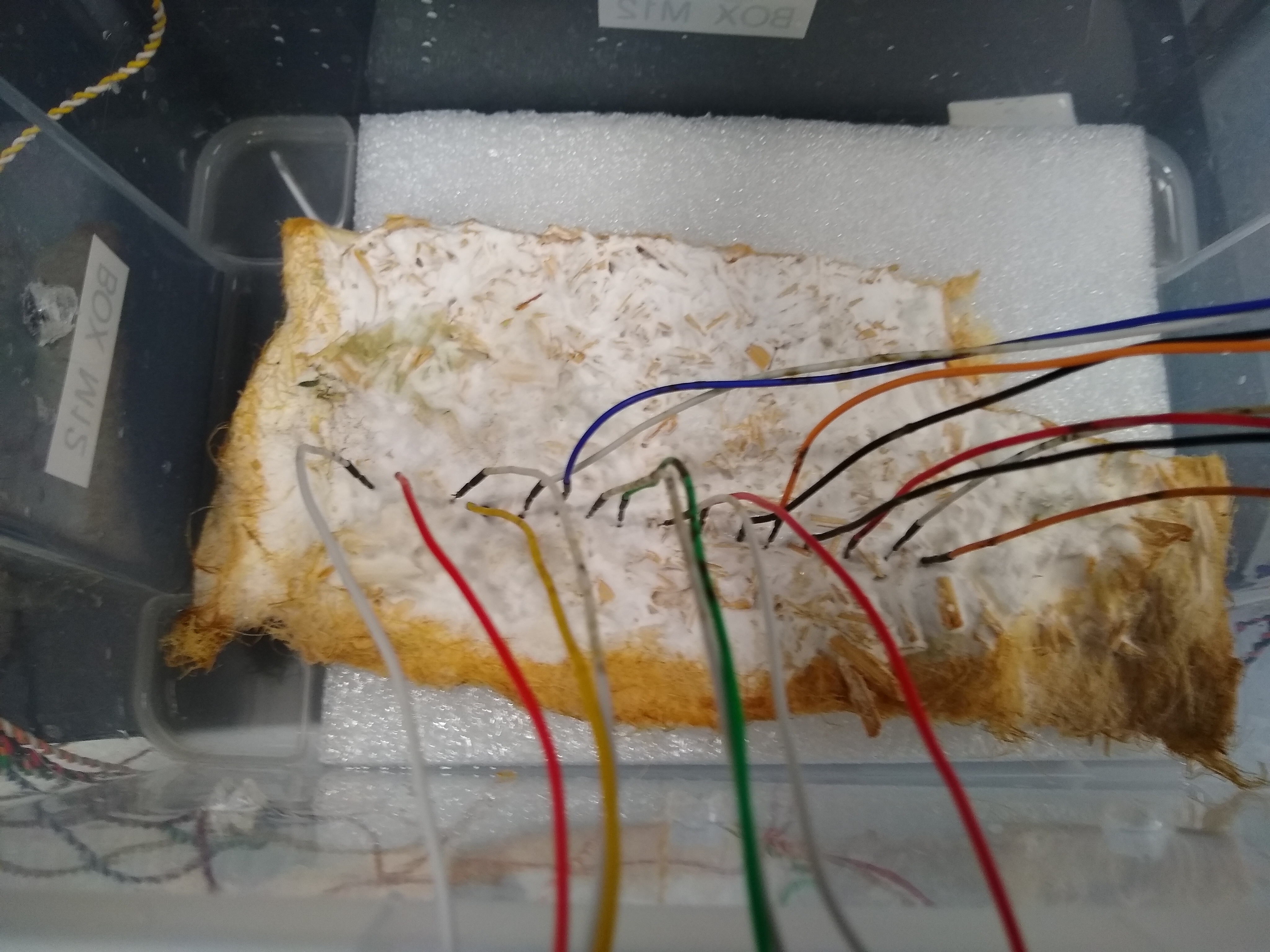}}\label{fig:electrodes}
\subfigure[]{\includegraphics[width=0.49\textwidth]{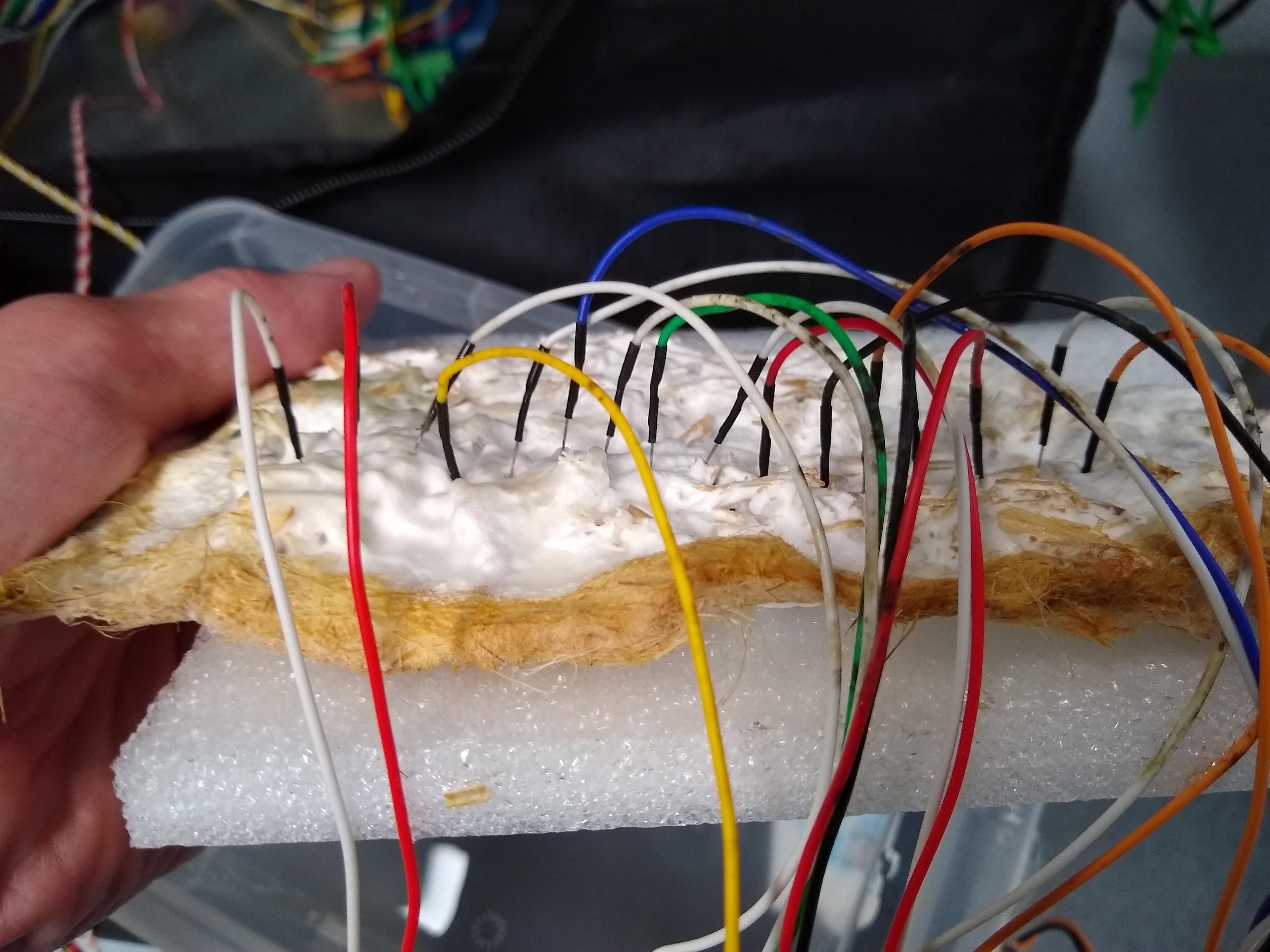}}\label{fig:electrodes1}
\subfigure[]{\includegraphics[width=0.49\textwidth]{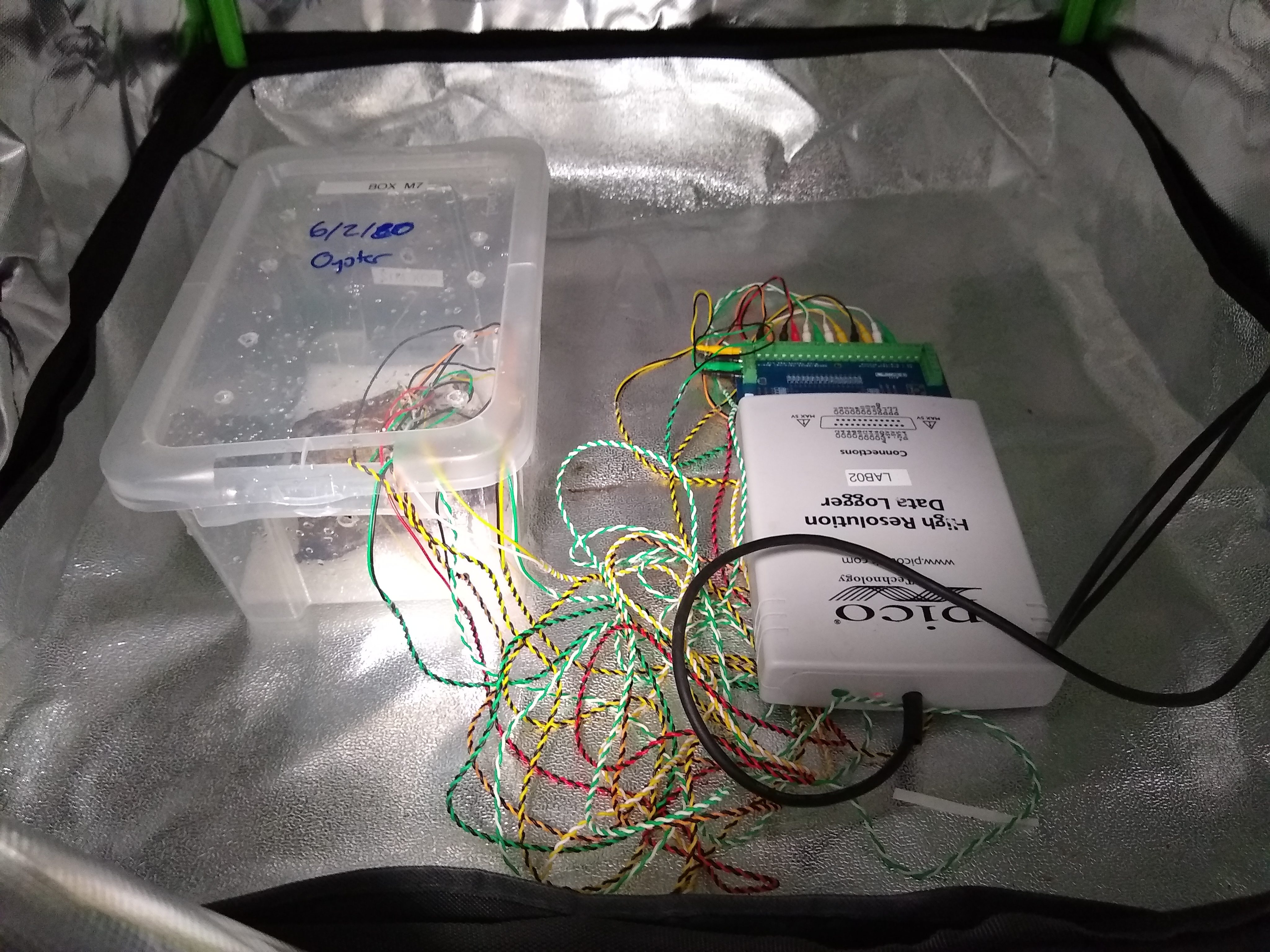}}\label{fig:tent}        
    \caption{Experimental setup. 
    (ab)~Exemplar locations of electrodes.
    (c)~Setup in the grow tent.
  }
    \label{fig:setup}
\end{figure}

A commercial strain of the fungus \emph{Pleurotus ostreatus} (collection code 21-18, Mogu S.r.l., Italy), previously selected for its superior fitness growing on the targeted substrate, was cultured on sterilised hemp shives contained in plastic (PP5) filter patch microboxes (SacO2, Belgium) that were kept in darkness at ambient room temperature c.~22\textsuperscript{o}C.
After one week of incubation, a hemp brick well colonised by the fungus was manually crumbled and spread on rectangular fragments, c. $12 \times12~{\text cm}^2$, of moisturised nonwoven hemp pads. When these fragments were colonised, as visualised by white and healthy mycelial growth on surface, they were used for experiments. 

Electrical activity of the colonised hemp pads was recorded using pairs of iridium-coated stainless steel sub-dermal needle electrodes (Spes Medica S.r.l., Italy), with twisted cables and  ADC-24 (Pico Technology, UK) high-resolution data logger with a 24-bit A/D converter. To keep electrodes stable we have been placing a polyurethane pad under the fabric. The electrodes were arranged in a line (Fig.~\ref{fig:electrodes}ab). The pairs of electrodes were pierced through the fabric and into the polyurethane pad. 

The fungal substrates pierced with electrodes was placed into 20~cm by 10~cm by 10~cm plastic boxes with tight lids.

We recorded electrical activity at one sample per second. During the recording, the logger has been doing as many measurements as possible (typically up to 600 per second) and saving the average value. We set the acquisition voltage range to 156~mV with an offset accuracy of 9~$\mu$V at 1~Hz to maintain a gain error of 0.1\%. Each electrode pair was considered independently with the noise-free resolution of 17 bits and conversion time of 60~ms. Each pair of electrodes, called channels, reported a difference of the electrical potential between the electrodes. Distance between electrodes was 1-2~cm. In each trial, we recorded eight electrode pairs, channels, simultaneously.   

To study the effect of chloroform we soaked a piece of filter paper c. 4~cm by 4~cm in chloroform (Sigma Aldrich, analytical standard) and placed the piece of paper inside the plastic container with the recorded fungal substrate.

The humidity of the fungal colonies was 70\%-80\% (MerlinLaser Protimeter, UK). The experiments were conducted in a room with ambient temperature 21\textsuperscript{o}C and in the darkness of protective growing tents (Fig.~\ref{fig:tent}).

We have conducted ten experiments, in each experiments we recorded electrical activity of the fungi via eight channels. 

\section{Results}
\label{results}

\begin{figure}[!tbp]
    \centering
    \includegraphics[width=0.8\textwidth]{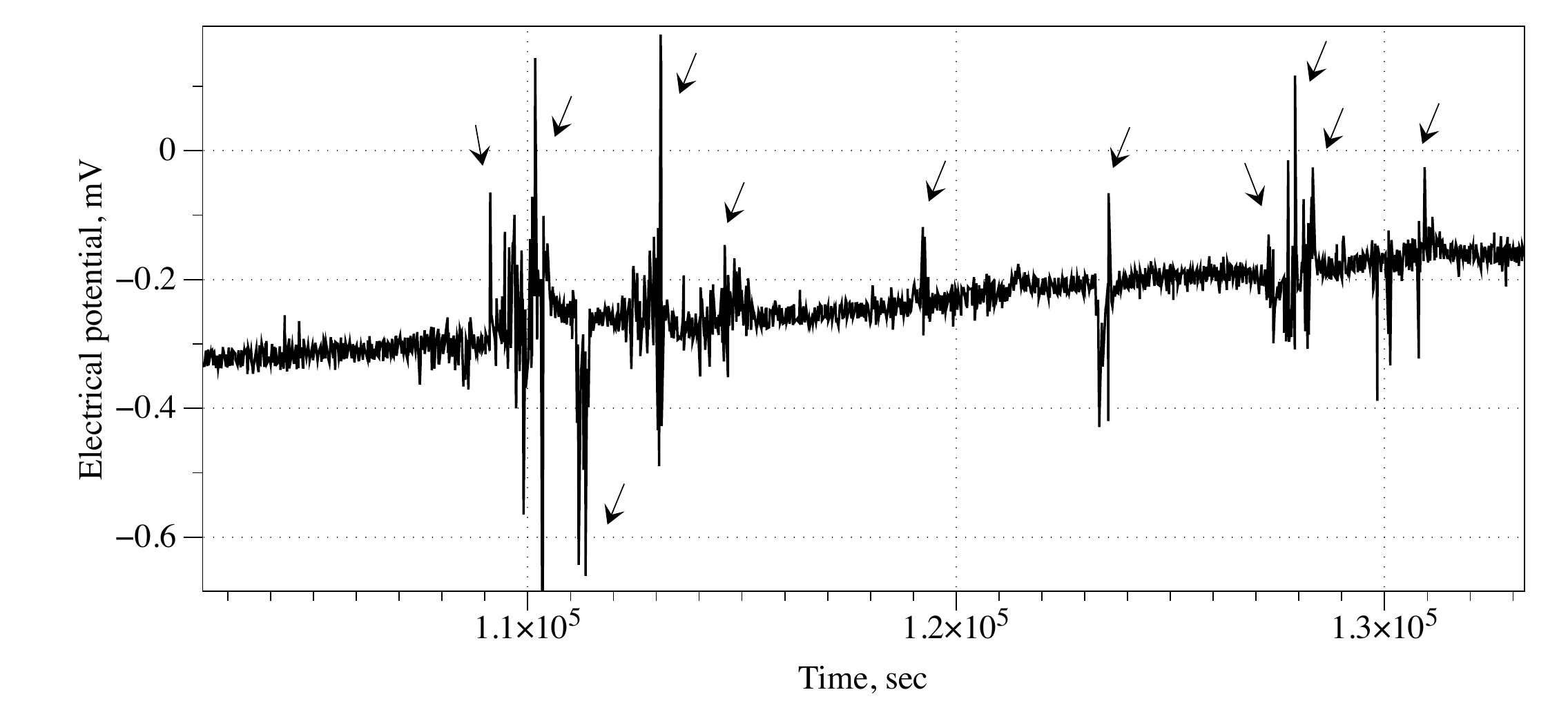}
    \caption{Example of spiking activity recorded from intact mycelium colonised hemp pad. Spikes are shown by arrows.}
    \label{fig:exampleSpiking}
\end{figure}

Mycelium colonised hemp pad exhibit patterns of electrical activity similar to that of spiking neural tissue. Examples of action potential like spikes, solitary and in trains,  are shown in Fig.~\ref{fig:exampleSpiking}.

\begin{figure}[!tbp]
    \centering
    \includegraphics[width=0.99\textwidth]{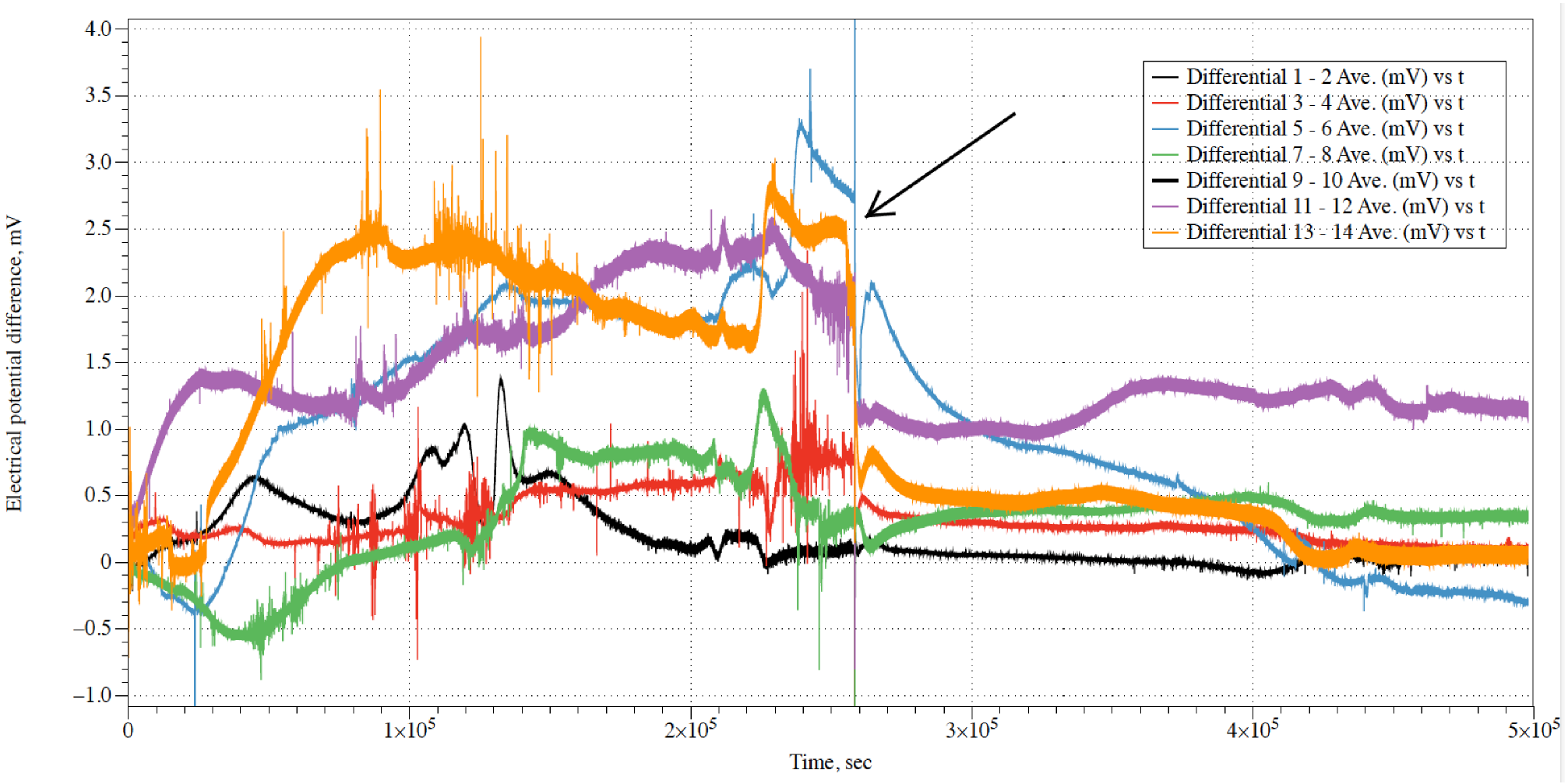}
    \caption{An example showing how the electrical activity of fungi changes when chloroform is introduced. The moment of the chloroform introduction is shown by arrow.}
    \label{fig:typicalresponse}
\end{figure}

Application of the chloroform to the container with fungi substantially affected the electrical activity of the fungi. An example of an extreme, i.e. where almost all electrical activity of mycelium seized, response is shown in Fig.~\ref{fig:typicalresponse}. In this example, the introduction of the chloroform leads to the suppression of the spiking activity and reduction of deviation in values of the electrical potential differences recorded on the channels. 

The intact mycelium composite shows median amplitude of the irregular movements of the baseline potential is 0.45~mV (average 0.64~mV, $\sigma=0.64$), median duration 29850~sec (average 67507~sec, $\sigma=29850$). After exposure to chloroform the baseline potential movements show median amplitude reduced to 0.16~mV (average 0.18~mV, $\sigma=0.12$) and median duration increased to 38507~sec (average 38114~sec, $\sigma=38507$). For the eight channels (pairs of differential electrodes) recorded exposure to chloroform led to nearly three times decrease in amplitude of the drifts of baseline potential and nearly 1.3 increase in duration of the drifts. Before exposure to chloroform the mycelium composite produced fast (i.e. less than 10-20 mins) spikes. Median amplitude of the spikes was 0.48~mV (average 0.52~mV, $\sigma=0.2$). Median duration of spikes was 62~sec (average 63~ec, $\sigma=18$), median distance between the spike 214~sec (average 189~sec, $\sigma=90$). After exposure to chloroform the mycelium composite did not show any spiking activity above level of background noise, which was for this particular recording c. 0.05~mV.

\begin{figure}[!tbp]
    \centering
    \includegraphics[width=0.99\textwidth]{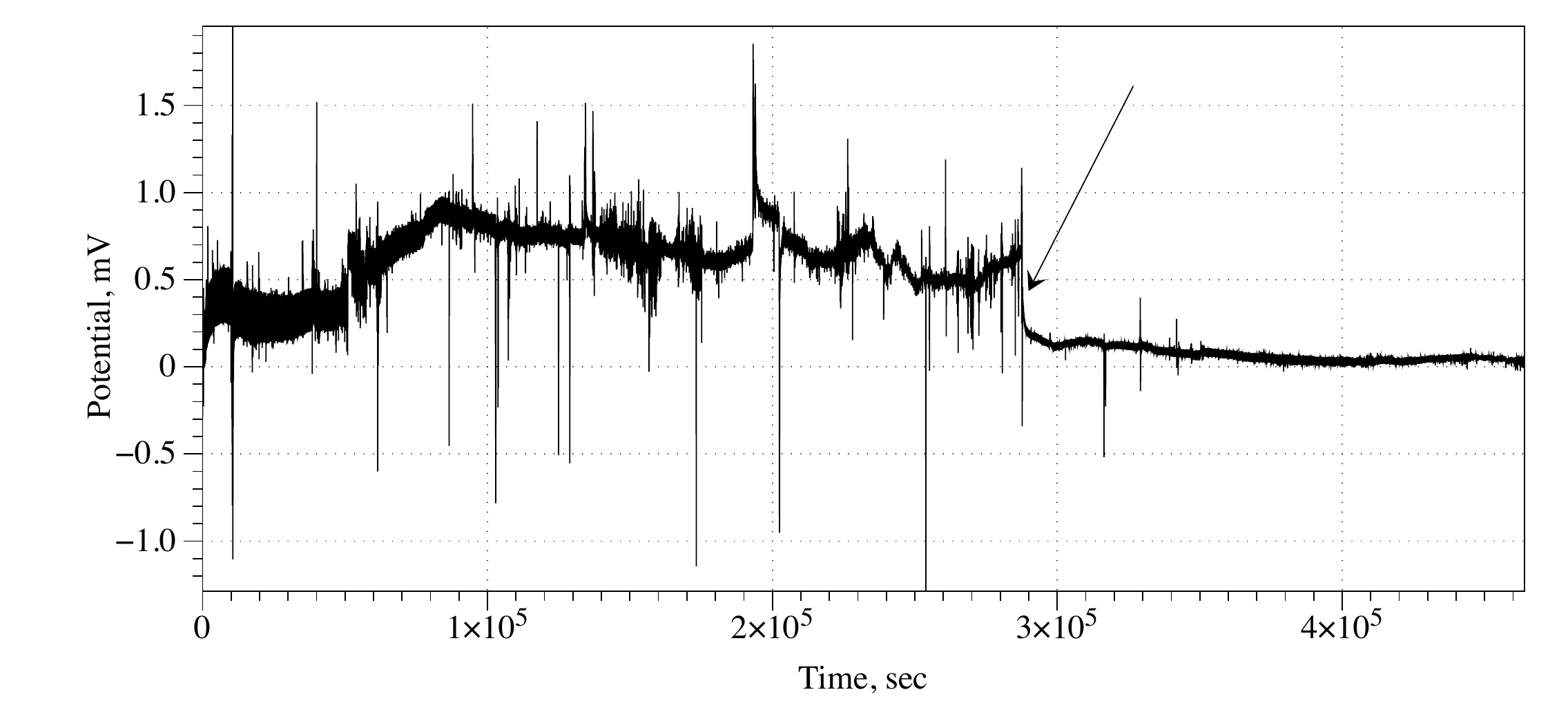}
    \caption{Example of reduced frequency of spiking under effect of chloroform vapour. Moment when a source of chloroform vapour was added into the container is shown by arrow. }
    \label{fig:ExampleReducedSpiking}
\end{figure}

In some cases the spiking activity is diminished gradually with decreased frequency and lowered amplitude, as exemplified in Fig.~\ref{fig:ExampleReducedSpiking}. Typically, the intact spiking frequency is a spike per 70~min while after inhalation of chloroform a spike per 254~min in the first 40-50~hours and decreased to nearly zero after. The median amplitude of intact mycelium spikes is 0.51~mV, average 0.74~mV ($\sigma$=0.59). Anaesthetised mycelium shows, spikes with median amplitude 0.11~mV, average 0.2~mV ($\sigma=0.2$). Spikes are not distributed uniformly but gathered in trains of spikes. In the intact mycelium there is a media of 3 spikes in the train, average number of spikes is 4.2 ($\sigma=4.4$). Median duration of a spike train is  84~min, average 112~min ($\sigma=32$). Media interval between trains is 53~min, average 55~sec ($\sigma=29$). Anaesthetised mycelium emits trains with median number of 2 spikes, average 2.5 spikes, average 2.5 spikes ($\sigma=0.84$). A median duration of such trains is 29~min, average 51~min ($\sigma=22$). The trains appear much rarely than the trains in the intact mycelium: median interval between trains is 227~min.

\begin{figure}[!tbp]
    \centering
    \includegraphics[width=1.1\textwidth]{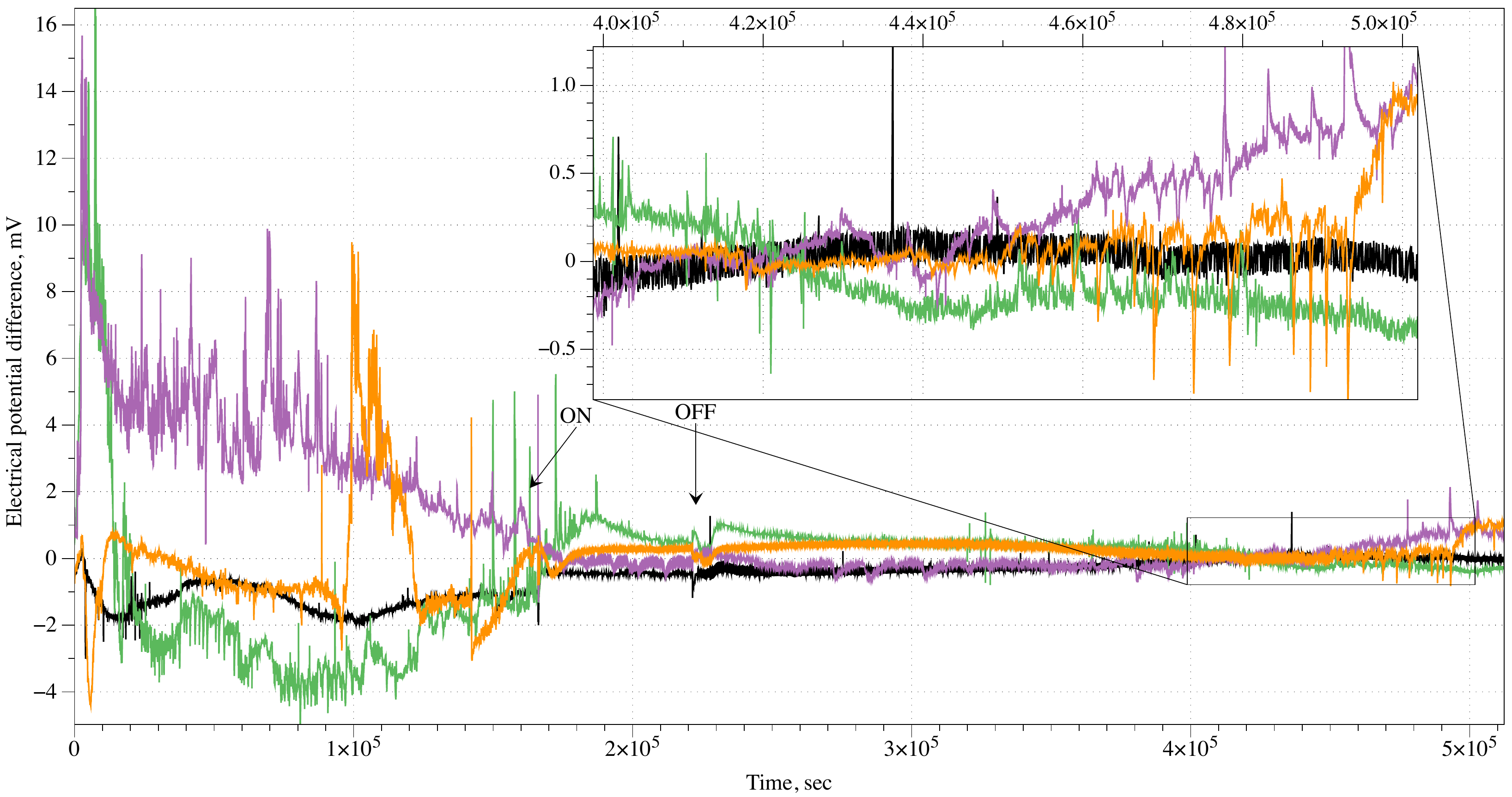}
    \caption{Example of electrical activity of mycelium colonised hemp pad before, during and after stimulation with chloroform vapour. Arrow labelled `ON' shows moment when a source of chloroform vapour was added was added into the enclosure, `OFF' when the source of chloroform was removed. The spiking activity of the mycelium recovering from anaesthesia is zoomed in.}
    \label{fig:recovery}
\end{figure}

In all ten but one experiment the container remained closed for over two-three days. By that time all kinds of electrical activity in mycelium bound substrate extinguished and the mycelium never recovered to a functional state. In experiment illustrated in Fig.~\ref{fig:recovery} we removed a source of chloroform after 16~hours and kept the container open and  well ventilated for an hour to remove any traces of chloroform from the air. 
The intact mycelium shows median frequency of spiking as one spike per 27~min, average 24~min. Median amplitude of the spikes is 3.4~mV, average 3.25~mV ($\sigma=1.45$). The anaesthetised mycelium demonstrates electrical spiking activity reduced in amplitude: median amplitude of spikes is 0.24~mV, average 0.32`mV ($\sigma=0.2$), and low frequency of spiking: median distance between spikes is 38~min, average 40~min. Electrical activity of the mycelium restores to above noise level c.~60~hours after the source of the chloroform is removed from the enclosure (insert in Fig.~\ref{fig:recovery}). Frequency of spikes is one spike per 82~min (median), average 88~min. The amplitudes of recovering spikes are 0.96~mV in median (average 0.93~mV, $\sigma=0.08$) which are three times less than of the spikes in the mycelium before the narcosis but nearly five times higher than of the spike of the anaesthetised mycelium.

\section{Discussion}
\label{discussion}

\begin{figure}[!tbp]
    \centering
   \subfigure[]{\includegraphics[width=0.7\textwidth]{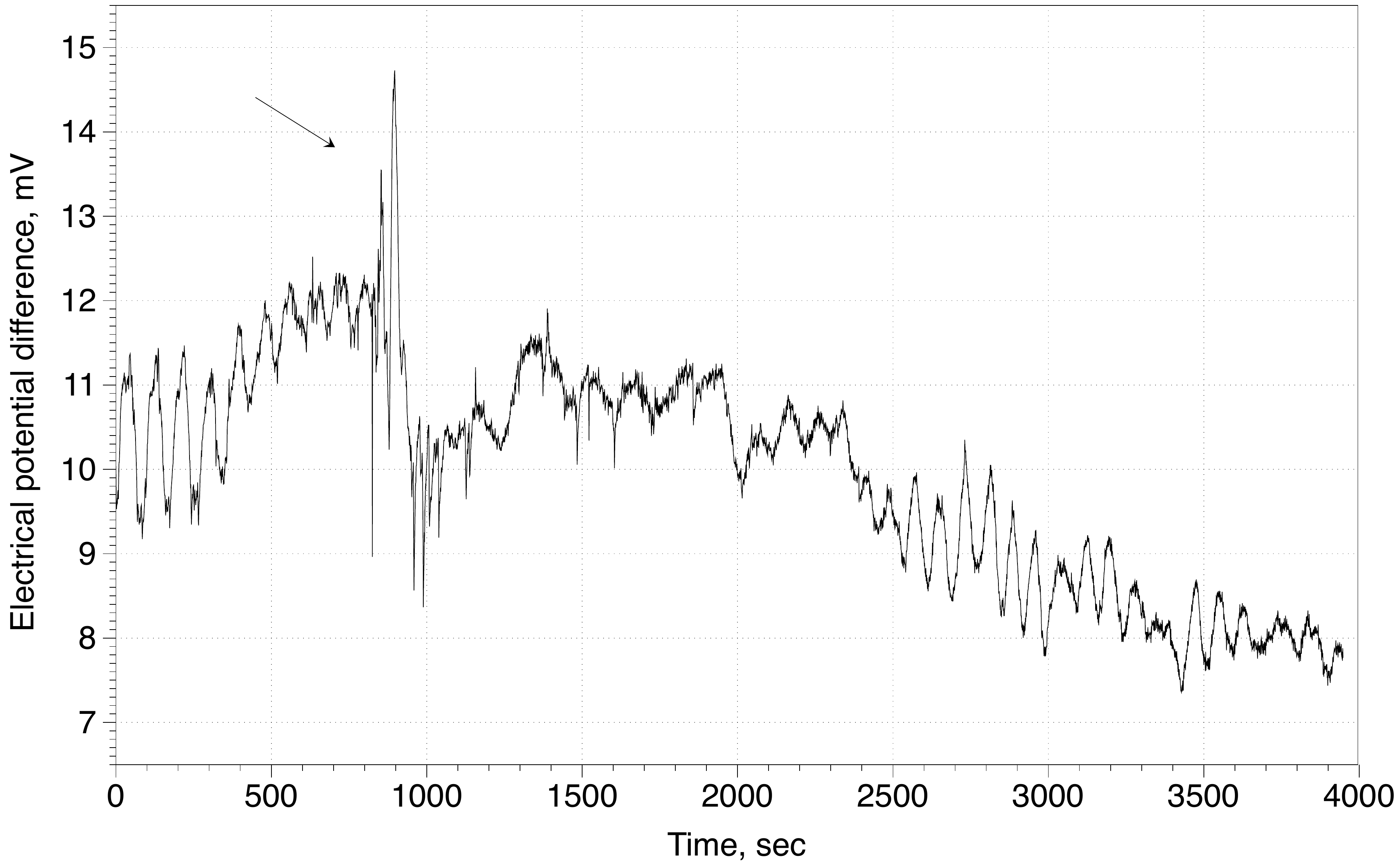}}
   \subfigure[]{\includegraphics[width=0.7\textwidth]{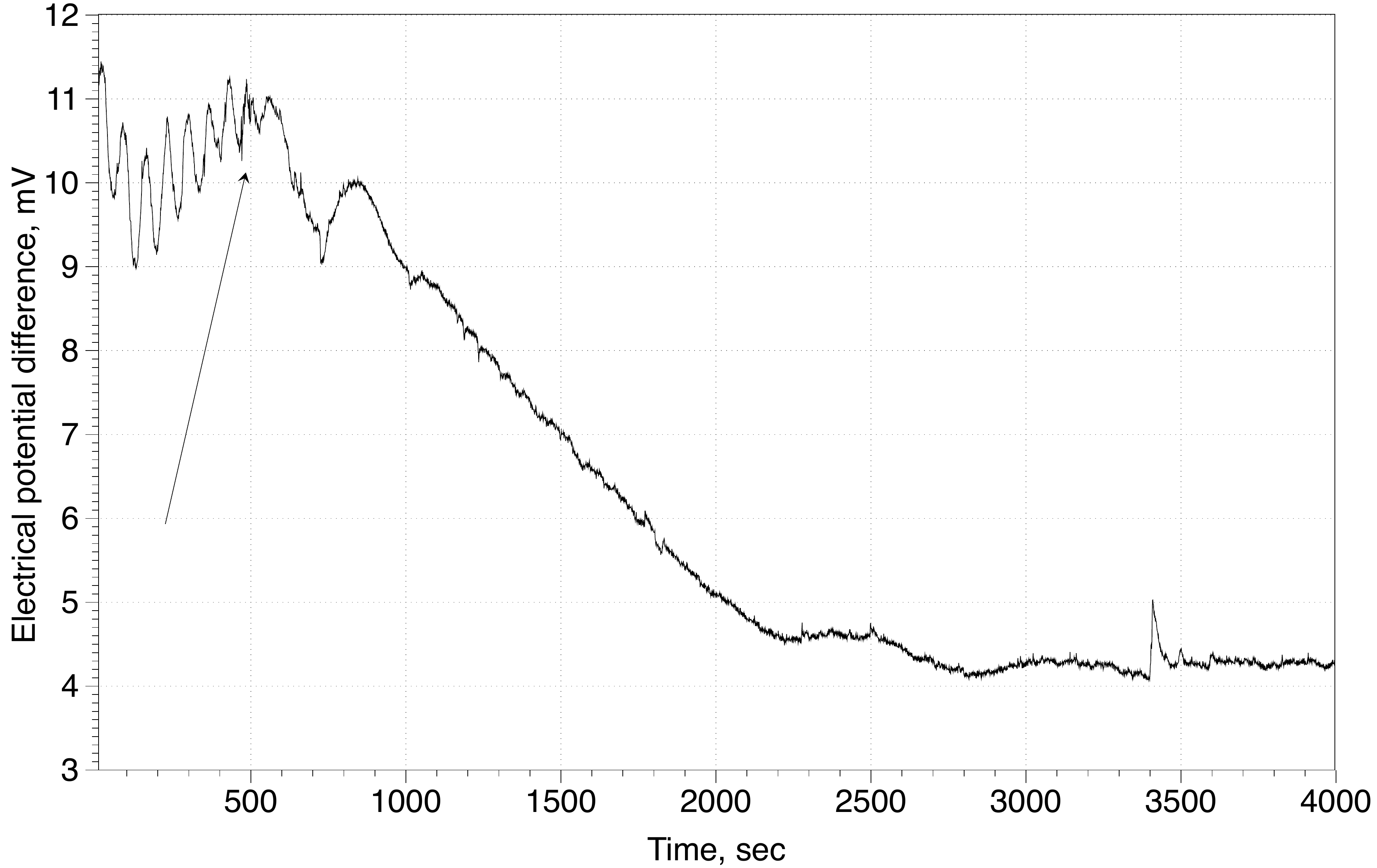}}
    \caption{Response of slime mould \emph{Physarum polycephalum} to trifluoroethane (Sigma Aldrich, UK). Electrical potential difference between two sites of 10~mm long protoplasmic tube was measured using aluminium electrodes, amplified and digitised with ADC-20 (Pico Technology, UK). 
    (a)~5~$\mu$L of trifluoroethane applied to a 5~mm$\times$5~mm piece of filter paper placed in the Petri dish with slime mould. 
    (b)~25~$\mu$L applied. Arrows indicate moments when the piece of paper soaked in trifluoroethane was placed in a Petri dish with the slime mould. }
    \label{fig:Physarum}
\end{figure}

We demonstrated that the electrical activity of the fungus \emph{Pleurotus ostreatus} is a reliable indicator of the fungi anaesthesia. When exposed to a chloroform vapour the mycelium reduces frequency and amplitude of its spiking and, in most cases, cease to produce any electrical activity exceeding the noise level. When the chloroform vapour is eliminated from the mycelium enclosure the mycelium electrical activity restores to a level similar to that before anaesthesia. The fungal responses to chloroform are similar to that recorded by us with slime mould \emph{Physarum polycephalum} (unpublished results). A small concentration of anaesthetic leads to reduced frequency and amplitude of electrical potential oscillation spikes of the slime mould, and some irregularity of the electrical potential spikes (Fig.~\ref{fig:Physarum}a). Large amount of anaesthetic causes the electrical activity to cease completely and never recover. 

With regards to directions of future research, as far as we are aware, the present paper is the first in the field, and therefore it rather initiates the research than brings any closure or conclusions. We know that anaesthetics block electrical activity of fungi (as well as slime moulds) however we do not know exact biophysical mechanisms of these actions. The study of biophysics and molecular biology of fungi anaesthesia could be a scope for future research. Another direction of studies could be the analysis of the decision making abilities of fungi under influence of anaesthetics. An experiment could be constructed when fungal hyphae are searching for an optimal path in a labyrinth when subjected to increasing doses of chloroform vapour. There may be an opportunity to make a mapping from concentrations of anaesthetic to geometry of the mycelium search path.

\section*{Acknowledgement}

This project has received funding from the European Union's Horizon 2020 research and innovation programme FET OPEN ``Challenging current thinking'' under grant agreement No 858132. The authors would like to acknowledge the collaboration of Mogu S.r.l. providing the living materials used in the experiments. All the experiments have been conducted in the Unconventional Computing Lab, UWE, Bristol. 


\end{document}